\documentstyle[twocolumn,prl,aps,epsf]{revtex}
\begin{document}
\draft
\title{Delayed feedback control of periodic orbits in
autonomous systems}
\author{Wolfram Just\thanks{e--mail:
wolfram@mpipks-dresden.mpg.de}}
\address{Max Planck Institute for Physics of Complex Systems,
N\"othnitzer Stra\ss e 38, D--01187 Dresden, Germany}
\author{Dirk Reckwerth, Johannes M\"ockel, 
Ekkehard Reibold\thanks{e--mail: reibold@exp1.fkp.physik.tu-darmstadt.de},
and 
Hartmut Benner\thanks{e--mail: benner@hrzpub.tu-darmstadt.de}}
\address{Institut f\"ur Festk\"orperphysik, Technische Universit\"at Darmstadt,
Hochschulstra\ss e 6, D--64289  Darmstadt, Germany}
\date{February 23, 1998}
\maketitle
\begin{abstract}
For controlling periodic orbits with delayed feedback methods the periodicity
has to be known a priori. We propose a simple scheme, how to detect
the period of orbits from properties of the control signal,
at least if a periodic but nonvanishing signal is observed. We
analytically derive a simple expression relating the delay,
the control amplitude, and the unknown period.
Thus, the latter can be computed 
from experimentally accessible quantities. Our
findings are confirmed by numerical simulations and electronic circuit
experiments.
\end{abstract}
\pacs{PACS numbers: 05.45.+b, 02.30.Ks, 07.50.Ek}
\newcommand{\be}{\begin{equation}\label}
\newcommand{\ee}{\end{equation}}
\newcommand{\bea}{\begin{eqnarray}\label}
\newcommand{\eea}{\end{eqnarray}}
\newcommand{\vx}{\mbox{\boldmath$x$}}
\newcommand{\vF}{\mbox{\boldmath$F$}}
\newcommand{\vz}{\mbox{\boldmath$z$}}
\newcommand{\vu}{\mbox{\boldmath$u$}}
\newcommand{\vv}{\mbox{\boldmath$v$}}
\newcommand{\vxi}{\mbox{\boldmath$\xi$}}
\newcommand{\veta}{\mbox{\boldmath$\eta$}}
\newcommand{\vzeta}{\mbox{\boldmath$\zeta$}}

Control techniques using time--delayed output signals are a very well
established field and known for at least half a century in the engineering 
and mathematical context (e.g.~\cite{bell} and references therein). 
Delayed feedback
control methods, which have for the physicists' purpose
been rediscovered in \cite{pyr},
are very useful since neither special knowledge of the system 
under consideration nor sophisticated reconstruction techniques are 
required, and the method is easily implemented in experiments \cite{exp}.
As a certain kind of drawback, the success of delayed feedback methods is
difficult to predict, and the stability analysis of the corresponding
delay systems shows a rich behaviour (e.g.~\cite{BS}). Only recently 
some progress in the understanding of general features has been made
in the physical context \cite{JRB}. Since control of actual periodic orbits 
with delayed feedback methods requires a delay time which is an integer 
multiple of the period, one runs into principle difficulties whenever the
period is not known a priori. Some empirical schemes have been
reported to circumvent such problems \cite{KPP}. They work quite well 
for special cases but no theoretical foundation has been proposed.
Here we address the problem that the period of the unstable
periodic orbit is unknown. A systematic strategy is developed
to obtain the desired period,
whenever a periodic control signal is observed. 

{\it Theoretical approach} -- To keep our approach as general as
possible the theoretical
considerations are based on a fairly arbitrary equation of motion
\be{aa}
\dot{\vx}=\vF(\vx(t),K(g[\vx(t)]-g[\vx(t-\tau)])) \quad .
\ee
Here $\vx$ denotes the phase space variables, 
$g[\vx]$ the measured scalar quantity,
$\tau$ the delay time, and $K$ the control amplitude. 
We do not specify the 
functional dependence of the systems on the control signal
$g[\vx(t)]-g[\vx(t-\tau)]$, since this dependence is in general
difficult to estimate from the experimental
point of view. Without control, $K=0$, 
the system should admit an unstable periodic orbit $\vxi(t)$
with period $T$ and Floquet exponent $\lambda+i\omega$, $\lambda>0$. 
We intend to stabilise this orbit.

Whenever the delay differs from the period, $\tau\neq T$, the orbit $\vxi$ 
does not yield a solution of the system subjected to control. However, 
the system admits a periodic solution $\veta$ with period $\Theta$.
Such a statement can even be proven rigorously \cite{Hale} provided that 
the delay mismatch $\tau-T$ is not too large.
In addition, the fictitious solution
$\veta$ tends towards the unstable orbit $\vxi$ in the limit 
$\tau\rightarrow T$. Of course, the period
of this fictitious orbit depends on the parameters of the system,
in particular on the delay time and the control amplitude, 
$\Theta=\Theta(K,\tau)$. We remind the reader, that the quantity $\Theta$
can be observed from the period of the control signal,
whenever the orbit $\veta$ is stable. In what follows we assume
that the system parameters are adjusted in such a way, 
i.~e.~we can observe the period $\Theta$ for different values of the
control amplitude $K$ and the delay time $\tau$. 

The strategy for the determination of the desired period $T$ is quite simple.
Since the orbit $\vxi$ yields a periodic orbit of the controlled system for
$\tau=T$, the measured period of the control signal obeys $\Theta(K,T)=T$.
Hence we simply have to look for zeros of the function $\Theta(K,\tau)-\tau$.
The latter can be measured in principle, 
provided we meet the assumption made
above. Nevertheless, it would be helpful if some analytical result
about the dependence of $\Theta$ on the delay and the control amplitude would
be available. We show that
up to second order in the mismatch
$\tau-T$ the relation
\be{ab}
\Theta(K,\tau)= T+ \frac{K}{K-\kappa} (\tau-T) + {\cal O}((\tau-T)^2)
\ee
holds.
Here $\kappa$ denotes a system parameter which captures all the details
concerning the coupling of the control force to the system.
Since the parameters
$\tau$ and $K$ are adjustable in experiments and $\Theta$ is a
measurable quantity, the desired period can be computed from eq.(\ref{ab})
using two data points.

In order to derive expression (\ref{ab})
we rewrite eq.(\ref{aa}) for the periodic orbit $\veta$ 
in terms of the dimensionless time $s=t/\Theta$ as
\be{ac}
\veta'(s)=\Theta \vF(\veta(s),K(g[\veta(s)]-g[\veta(s-\tau/\Theta)])) 
\ee
and
\be{aca}
\veta(s)=\veta(s+1) \quad .
\ee
Since eq.(\ref{ab}) represents a Taylor expansion we are looking for the
derivative $\partial_{\tau} \Theta|_{\tau=T}$. For that reason one
takes the derivative of eq.(\ref{ac}) with respect to $\tau$, keeping in
mind that the periodic solution $\veta$ depends explicitly on $\tau$.
\bea{ad}
& & \left(\partial_{\tau} \veta\right)' -
\Theta D_1 \vF(\dots) \partial_{\tau}\veta(s) 
- \Theta K d_2\vF(\dots) \nonumber \\
&\cdot&
\left\{ D g[\veta(s)] \partial_{\tau} \veta(s)
- D g[\veta(s-\tau/\Theta)]
\partial_{\tau} \veta(s-\tau/\Theta)
\right\} \nonumber \\
&=& (\partial_\tau \Theta) \vF(\dots)
+ \Theta K d_2 \vF(\dots) \nonumber\\
&\cdot&
\left\{D g[\veta(s-\tau/\Theta)] 
\veta'(s-\tau/\Theta) \right\}
\partial_ \tau (\tau/\Theta) \quad .
\eea
Here $D_1$ and $d_2$ denote the derivative with respect to the first/second
argument of $\vF$, and the arguments abbreviated by $\cdots$ coincide with 
those from eq.(\ref{ac}). The contributions involving the derivative
of the orbit with respect to the explicit $\tau$--dependence,
$\partial_{\tau}\veta$, have been collected on the left hand side.
The boundary value problem (\ref{ad}), (\ref{aca}) 
determines both, $\partial_{\tau} \Theta$ as well as 
$\partial_{\tau} \veta$. In order to separate the former quantity we trace back
to the fact that the linear operator on the left hand side of
eq.(\ref{ad}) admits a vanishing eigenvalue. The corresponding
Goldstone mode is related to the translation invariance in time of the
original system. In fact,
taking the derivative of eq.(\ref{ac}) with respect to $s$ one obtains
\bea{ae}
0&=& \left(\veta'\right)' - \Theta D_1 \vF(\dots)
\veta'(s) - \Theta K d_2\vF(\dots)  \nonumber \\
&\cdot& \left\{ 
Dg[\veta(s)] \veta'(s) - Dg[\veta(s-\tau/\Theta)] \veta'(s-\tau/\Theta)
\right\} \quad .
\eea
Eq.(\ref{ae}) just states that $\veta'$ yields the
right--\-null\-ei\-gen\-function. 
Within the canonical scalar product
$\int_0^1 \vv(s) \vu(s) ds$ we denote the corresponding
left--null\-ei\-gen\-function by $\vzeta(s)$. 
All the terms on the left hand side of eq.(\ref{ad}), which
involve $\partial_{\tau} \eta$ vanish identically after multiplication
with $\zeta$. Hence we are left with
\bea{af}
0 &=&
\partial_\tau \Theta  \int_0^1 \vzeta(s) \vF(\dots) \, ds
+ \Theta K \partial_ \tau (\tau/\Theta) \nonumber\\
&\cdot& \int_0^1 \vzeta(s) d_2 \vF(\dots) 
\left\{D g[\veta(s-\tau/\Theta)] 
\veta'(s-\tau/\Theta) \right\} \, ds
\eea
The details of the system, which are only contained in the integrals,
are now condensed to simple numbers. But in general the integrals depend
on the delay $\tau$ and in particular on the control amplitude $K$ through 
the left--eigenfunction $\vzeta$ (cf.~eq.(\ref{ae})). For that reason we
evaluate eq.(\ref{af}) at $\tau=T$. Then $\Theta=\tau$ holds and the delay
in the arguments of $\veta$ drops by virtue of the boundary condition
(\ref{aca}). Due to the same argument the linear operator (\ref{ae}) and
therefore the eigenfunction $\vzeta$ becomes independent of $K$. Hence
the integrals become constant real numbers and eq.(\ref{af}) yields
\be{ag}
0= \kappa \partial_\tau \Theta|_{\tau=T} 
+ T K \partial_ \tau (\tau/\Theta)|_{\tau=T}
\ee
Here $\kappa$ denotes the ratio of the integrals occurring in eq.(\ref{af}).
We solve for $\partial_\tau \Theta|_{\tau=T}$, and obtain eq.(\ref{ab})
from a simple Taylor series expansion.

{\it Numerical simulations} -- We demonstrate the
applicability 
of our analytical results by
numerical simulations in an autonomous system. First of all stabilisation
of periodic orbits by delay methods requires a finite torsion, 
i.~e.~a finite frequency in the Floquet exponent of the controlled
orbit \cite{JRB}. Since autonomous equations always 
admit a vanishing exponent 
a finite frequency can be realised in dissipative three dimensional 
models only by a complete flip of the 
neighbourhood of the orbit. For that reason certain equations like
the Lorenz model cannot be stabilised at all by delay methods, apart
from the fixed points for which the reasoning given above does not
apply. Therefore we concentrate here on the R\"ossler equations
as a certain minimal model for our purpose.
\bea{ah}
\dot{x}_1 &=& -x_2 -x_3 -K\left(g[\vx(t)]-g[\vx(t-\tau)]\right)\nonumber\\
\dot{x}_2 &=& x_1 + a x_2 -K\left(g[\vx(t)]-g[\vx(t-\tau)]\right)\nonumber\\
\dot{x}_3 &=& b+ x_1 x_3 - c x_3
\eea
Our results do not seem to depend significantly on the coupling of the 
control force to the original equations of motion and
on the particular choice of the scalar quantity $g[\vx]$. We have used a 
bounded quantity in order to avoid diverging solutions. The results 
presented here correspond to the choice $g[\vx]=\tanh[(x_1+x_2)/10]$. 
In addition, the 
system parameters have been fixed to the values $a=b=0.2$, $c=5.7$ to
ensure chaotic dynamics in the absence of control. 
For our control purpose we concentrate on the period--two orbit in the
canonical Poincare map with $T=11.758\ldots$, $\lambda T=1.256\ldots$, and
$\omega T=\pi$. Numerical simulations have been performed by means of an
adaptive stepsize Runge--Kutta method of order four, together with a
cubic spline for the delay from the NRF library \cite{nrf}. 

\begin{figure}
\begin{center}\mbox{\epsffile{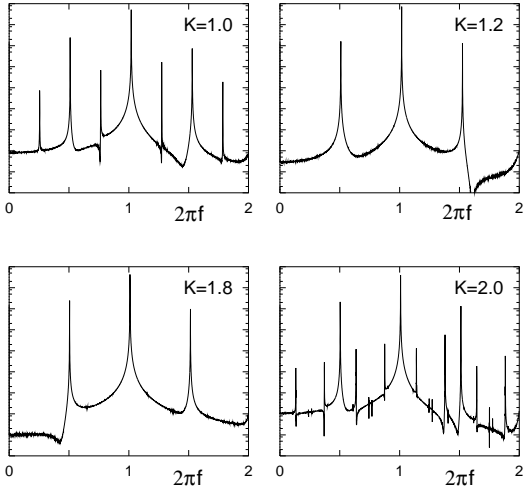}}\end{center}
\caption[ ]{Absolute square value of the Fourier transform of the scalar 
quantity $g[\vx]$ for $\tau-T=1.0$ 
in the vicinity of the lower and upper
stability threshold. The transform has been performed for a series of length
$1024\times \tau$ discarding a transient of $100\times \tau$. The
spectrum has not been properly normalised and the abscissa extends over
9 decades. \label{figa}}
\end{figure}

\begin{figure}
\begin{center}\mbox{\epsffile{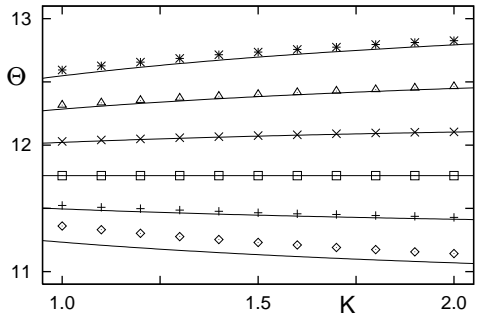}}\end{center}
\caption[ ]{Dependence of the period $\Theta$ on the control amplitude for
various delay times $\tau$, from bottom to top $\tau-T=$ $-1.0$, $-0.5$,
$0.0$, $0.5$, $1.0$, $1.5$: numerical
simulations (symbols) and analytical
expression (\ref{ab}) with $\kappa=-0.9$ (lines).
For a few data points outside the stabilised regime,
the period was estimated
from a dominant peak in the power spectrum.\label{figb}}
\end{figure}

For a quite large range of delay times $\tau$
one observes two critical values of the control amplitude which limit
an interval where a stable periodic orbit $\veta$ can be observed.
From the Fourier transform of the scalar quantity $g[\vx]$ it is evident 
(cf.~fig.\ref{figa}) that
at the lower critical value the orbit loses stability via a flip
bifurcation, whereas at the upper critical value a Hopf bifurcation occurs.
In order to check the accuracy of eq.(\ref{ab}),
the period $\Theta$ of the fictitious orbit has been
extracted from the peaks in the Fourier spectra of the control 
signal. The dependence of $\Theta$
on the control amplitude for several delay times
is summarised in fig.\ref{figb} and compared with our
analytical expression.
The apparent systematic deviation of the analytical
result just comes from the fact that the latter is a first--order
approximation to the curved manifold $\Theta(K,T)$ in the three--dimensional
$K$--$T$--$\Theta$ space. In summary, eq.(\ref{ab}) describes
the observed periods quite accurately.

Finally we have checked, whether eq.(\ref{ab}) successfully
predicts the period of the unstable periodic orbit $\vxi$ whenever a few data
points are accessible. To this end
we evaluated the $K$--dependence of
the power spectrum of the control signal within a regime where a periodic
signal can be observed. 
Starting from $\tau=14.0$, which differs tremendously from the true period,
we evaluate $\Theta$ for $K=0.8$, $0.9$, and $1.0$ to obtain
$\kappa=-0.8\pm0.01$ and $T=11.745\pm0.015$
from eq.(\ref{ab}). The accuracy of $T$
is in fact of the order of the numerical resolution of the power spectra.
In that sense the result is striking.

\begin{figure}
\begin{center}\mbox{\epsffile{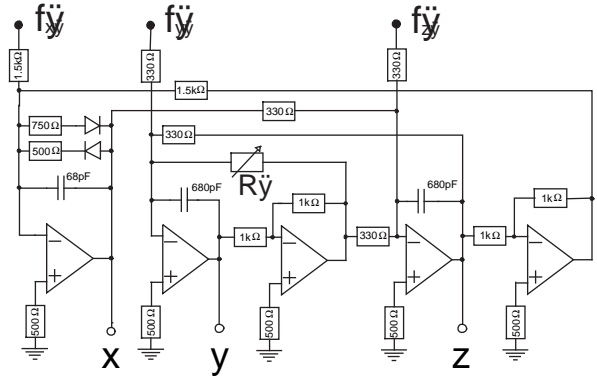}}\end{center}
\caption[ ]{Experimental setup of the nonlinear electronic circuit
without the time--delayed feedback device. Experiments have been performed
at $R=110 \Omega$.}
\label{nec}
\end{figure}

{\it Experiments} -- To illustrate the experimental 
accessibility of our analytical results
we have performed measurements on a nonlinear electronic circuit
(cf.~fig.\ref{nec}).
The circuit consists of several operational amplifiers (three acting as
integrators, two as inverters) with associated feedback components.
The nonlinearity is provided by the diodes. 
The voltages probed at $x,y,z$ can be considered as the degrees of freedom
in our experiment. At $f_{x}, f_{y}, f_{z}$ external signals can be 
fed into the
system for control purpose. Typical frequencies
of the circuit are about $600 \mbox{kHz}$.

Without control
the system undergoes a period--doubling cascade to 
chaos on variation of the resistance $R$, ending up in a 
R\"ossler--type attractor.
Topological analysis \cite{top}
of this three--dimensional system yielded a 
frequency of $\pi / T$ in the Floquet exponent for the unstable 
period--one orbit of the chaotic attractor. This corresponds
to a complete flip of the neighbourhood of this orbit.
Therefore the orbit is accessible to time--delayed feedback control.

The control device consists of a cascade of electronic 
delay lines with a limiting frequency of about $3 \mbox{MHz}$ 
and several operational amplifiers acting
as preamplifier, subtractor, or inverter. 
The device allows to apply a control force of the form 
$F(t) = - K [U(t) - U(t- \tau )]$
with $\tau$--range $10 \mbox{ns}  \ldots 21 \mu \mbox{s}$. 
Our feedback scheme consisted of coupling the voltage at $z$ via the
control device to $f_z$.

\begin{figure}
\begin{center}\mbox{\epsffile{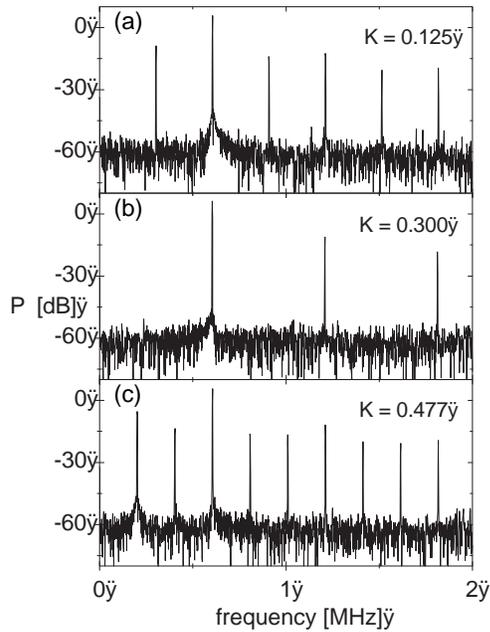}}\end{center}
\caption[ ]{Power spectrum of $x(t)$ for 
$\tau -T = 2.5\mbox{ns}$:
(a) below the lower, (b) between both, and (c) above the upper
stability threshold.}
\label{spec_exp}
\end{figure}

To check the coincidence with our analytical results we looked for periodic 
behaviour of our nonlinear circuit by
sweeping the control amplitude $K$ at fixed $\tau$.
By increasing $K$ the system undergoes an inverse period--doubling
cascade ending up in a period--one state. This periodic state
yields the desired value $\Theta$. A further increase of $K$
results in a Hopf bifurcation destroying the stability of the 
periodic state (cf.~fig.\ref{spec_exp}).

\begin{figure}
\begin{center}\mbox{\epsffile{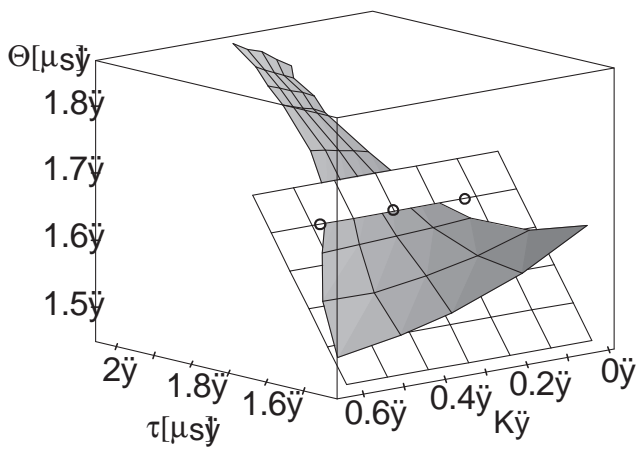}}\end{center}
\caption[ ]{Measured values of the period $\Theta$ depending on
$\tau$ and $K$. The white plane corresponds to $\Theta = \tau$.
The circles indicate the points where
spectra shown in fig.\ref{spec_exp} have been obtained.}
\label{surface}
\end{figure}

Figure \ref{spec_exp}(a) shows the main frequency 
at $605.5 \mbox{kHz}$ and its subharmonic at $302.7 \mbox{kHz}$ corresponding
to the flip bifurcation at the lower stability threshold.
At the upper stability threshold a Hopf bifurcation yields
an incommensurate frequency component at $201.4 \mbox{kHz}$. 
We checked that frequency locking did not occur.
By measuring $\Theta$ for various $\tau$ and $K$ values one obtains 
the greyshaded surface displayed in fig.\ref{surface}. 
Note that for correct delay time $\tau = T$ one automatically gets 
$\Theta = \tau = T$. Within an experimental error of $1\mbox{ns}$ the intersection with the surface $\Theta=\tau$
yields a straight line with $\tau = 1.656 \mu \mbox{s}$.
Since the curvature of the surface in direction of $\tau$ is
negligible for $\tau$ values close to the real period, i.~e.~$\pm 10\%$,
the coincidence with our analytical expression (\ref{ab}) is quite reasonable
in this region.
Calculation of the system parameter yields
$\kappa = -0.31 \pm 0.01$. For larger delay mismatch
eq.(\ref{ab}) can still be used iteratively in the sense of a
Newton method for detecting the exact period $T$.

{\it Conclusion} -- We have shown that the period of true periodic unstable 
orbits can be
obtained from the properties of the control signal, at least if a
periodic signal can be realised. Our approach is based on
the fact that the true periodic orbit of the uncontrolled system is
deformed into
a fictitious periodic orbit by the control if the delay time
differs from the true period. Our analytical expression (\ref{ab}) relates the
fictitious period $\Theta$ with the true period $T$, the delay $\tau$, and
the control amplitude $K$. Peculiarities of the system enter only 
through a single parameter $\kappa$. Of course, 
our result does not guarantee that the orbit becomes stable if
the delay time is adapted without changing the control 
amplitude (cf.~fig.\ref{surface}). 
However, in order to keep the fictitious orbit stable during such
an adaption process one may for example monitor the power spectrum of the 
control signal (cf.~fig.\ref{spec_exp}), since an instability is indicated
by the
occurrence of additional peaks in the spectrum. 

{\it Acknowledgement} -- 
This project of SFB 185 ''Nichtlineare Dynamik''  
was partly financed 
by special funds of the Deutsche Forschungsgemeinschaft.
We are indebted to F.~Laeri and M.~M\"uller for the use of their 
delayed feedback control device.

\end{document}